\documentclass[ 
  aip,
 amsmath,amssymb,
  reprint,
]{revtex4-1}

\usepackage{graphicx}
\usepackage{dcolumn}
\usepackage{bm}
\usepackage[normalem]{ulem}

\usepackage[utf8]{inputenc}
\usepackage[T1]{fontenc}
\usepackage{mathptmx}
\usepackage{etoolbox}
\usepackage{comment}

\usepackage{mathtools}

\usepackage{hyperref}
\hypersetup{
    colorlinks,
    citecolor=blue,
    filecolor=blue,
    linkcolor=blue,
    urlcolor=black
}

\newcommand\diff{\mathrm{d}}
\renewcommand\Re{\text{Re}}
\renewcommand\Im{\text{Im}}

\makeatletter
\def\@email#1#2{%
 \endgroup
 \patchcmd{\titleblock@produce}
  {\frontmatter@RRAPformat}
  {\frontmatter@RRAPformat{\produce@RRAP{*#1\href{mailto:#2}{#2}}}\frontmatter@RRAPformat}
  {}{}
}%
\makeatother
\begin{document}

\preprint{AIP/123-QED}

\title[Complexified Synchrony]{Complexified Synchrony}

\author{Seungjae Lee}
\thanks{Corresponding Author (shared)}
\email{seungjae.lee@tu-dresden.de}
\affiliation{Chair for Network Dynamics, Center for Advancing Electronics Dresden (cfaed) and Institut für Theoretische Physik, Technische Universität Dresden, 01062 Dresden, Germany}

\author{Lucas Braun}
\thanks{These authors contributed equally to this work}
\affiliation{Chair for Network Dynamics, Center for Advancing Electronics Dresden (cfaed) and Institut für Theoretische Physik, Technische Universität Dresden, 01062 Dresden, Germany}
\affiliation{Schülerforschungszentrum Südwürttemberg (SFZ), 88348 Bad Saulgau, Germany}
\affiliation{Gymnasium Wilhelmsdorf, Pfrunger Straße 4/2, 88271 Wilhelmsdorf, Germany}
\author{Frieder B\"onisch}
\thanks{These authors contributed equally to this work}
\affiliation{Chair for Network Dynamics, Center for Advancing Electronics Dresden (cfaed) and Institut für Theoretische Physik, Technische Universität Dresden, 01062 Dresden, Germany}
\author{Malte Schr\"oder}
\affiliation{Chair for Network Dynamics, Center for Advancing Electronics Dresden (cfaed) and Institut für Theoretische Physik, Technische Universität Dresden, 01062 Dresden, Germany}
\author{Moritz Th\"umler}
\affiliation{Chair for Network Dynamics, Center for Advancing Electronics Dresden (cfaed) and Institut für Theoretische Physik, Technische Universität Dresden, 01062 Dresden, Germany}
\author{Marc Timme}
\thanks{Corresponding Author (shared)}
\email{marc.timme@tu-dresden.de}
\affiliation{Chair for Network Dynamics, Center for Advancing Electronics Dresden (cfaed) and Institut für Theoretische Physik, Technische Universität Dresden, 01062 Dresden, Germany}
\affiliation{Cluster of Excellence Physics of Life, TU Dresden, 01062 Dresden, Germany
}
\affiliation{Lakeside Labs, 9020 Klagenfurt, Austria} 

\date{\today}

\begin{abstract}
The Kuramoto model and its generalizations have been broadly employed to characterize and mechanistically understand various collective dynamical phenomena, especially the emergence of synchrony among coupled oscillators. Despite almost five decades of research, many  questions remain open, in particular for finite-size systems. Here, we generalize recent work [Phys. Rev. Lett. \textbf{130}, 187201 (2023)] on the finite-size Kuramoto model with its state variables analytically continued to the complex domain and also complexify its system parameters. Intriguingly, systems of two units with purely imaginary coupling do not actively synchronize even for arbitrarily large magnitudes of the coupling strengths, $|K| \rightarrow \infty$, but exhibit conservative dynamics with asynchronous rotations or librations for all $|K|$.
For generic complex coupling, both, traditional phase-locked states and asynchronous states generalize to complex locked states, fixed points off the real subspace that exist even for arbitrarily weak coupling. We analyze a new collective mode of rotations exhibiting finite, yet arbitrarily large rotation numbers. Numerical simulations for large networks indicate a novel form of discontinuous phase transition. We close by pointing to a range of exciting questions for future research.
\end{abstract}

\maketitle

\begin{quotation}Synchronization, the emergence of the temporal order of several state variables, has been ubiquitously observed across natural and engineered networks of interacting units, from neural circuits in the brain to electric power grids. In its simplest instances, all variables of a system become identical over time. Other forms of synchrony include achieving one common frequency shared by all units (frequency synchronization) and achieving fixed phase-differences (phase-locking). In the past 70 years, analytic continuation of system variables and parameters into the complex domain was successful in revealing and theoretically establishing core aspects of statistical physics, fractal geometry, quantum mechanics and other fields. In this article, we complexify coupled dynamical systems and explore the impact of analytically continuing state variables and parameters of the Kuramoto model, a fundamental model for investigating synchronization. We find that complex synchrony, a locking phenomenon represented by fixed points in the complex domain, is a generic feature of these complexified models. At the same time, real-variable synchronization seems to be the exception. For instance, systems with purely imaginary coupling strengths $K$ do not exhibit attractive locked states even if the coupling is arbitrarily strong, $|K|\rightarrow \infty$. We numerically and analytically highlight further features of complexified Kuramoto models, including analogies to phase-shifted and delay-coupled systems as well as new phase transitions at low $|K|$ to strong forms of synchrony.
\end{quotation}

\section{\label{sec:int} Introduction}
The Kuramoto model and its generalizations characterize the dynamics of a wide range of natural and engineered systems, from fireflies to neural circuits and from Josephson junctions to power grids~\cite{Ermentrout1991,josep1,josep2,josep3,josep4,josep5,WS-original2,Nishikawa_2015,Witthaut2022,Filatrella2008}. The original model was proposed in 1975 as a simple model for understanding the emergence of synchrony among coupled oscillators~\cite{Kuramoto1975,Strogatz2000,kuramoto2003chemical}. It has found ubiquitous applications across many areas in physics, biology, engineering and beyond~\cite{Acebron2005,RODRIGUES20161,DORFLER20141539}. 

With its many generalizations, including the Kuramoto-Sakaguchi model~\cite{sakaguchi1,OMELCHENKO201374}, the Winfree model~\cite{winfree1,winfree2,winfree3} and the Lohe model or higher-dimensional Kuramoto models~\cite{Lohe_WS,Lohe_2009,barioni1,barioni2,g_KM1,g_KM2,g_KM4,g_KM5,g_KM6,g_KM7, witthaut2014kuramoto, witthaut2017classical, Witthaut2022, Lee_2023,Lee_2024,PRX_generalized,Crnkic_2024}, it describes various synchronization and desynchronization phenomena, first and second order phase transitions to synchrony~\cite{pazo2005,pietras2018,Hu2014}, the coexistence of synchrony with other states and chimera states featuring both some variables in synchrony and others with asynchronous irregular dynamics~\cite{kuramoto2002,timme2006topology,abrams_chimera2008,abrams2004,Panaggio_2015,Omelchenko_2013}.

Most results on the Kuramoto models have been obtained in the thermodynamic limit $N \rightarrow \infty$, where phase transitions occur in the form of bifurcations of lower-dimensional descriptions~\cite{OA1,OA2,OA3,mobius_WS,Strogatz1991}. However, for finite-$N$ models despite them more closely describing reality, most results are purely numerical, sometimes employing intricate technical approaches~\cite{Peter2018,Peter2019} and analytical approaches that may offer stronger mechanistic insights are largely missing to date.
Recent work~\cite{Moritz_thuemler2023} has studied the Kuramoto model analytically continued to complex state variables, yet still with all parameters staying real, in part to access finite-$N$ systems mathematically. The specific work uncovered a new form of synchrony and especially highlighted the option of understanding finite systems via analytically accessing complex locked states, fixed points that exist off the real invariant subspace in the complex domain even for small $K$. The complex locked states represent generalized forms of synchrony reached even if the coupling is not strong enough to induce synchrony in the real-variable system. In contrast to previous studies on finite systems, the existence of these fixed points enables analytic access to the dynamical states for finite systems without requiring standard phase-locking (and thus sufficiently strong coupling).

In the past, analytic continuation has been highly successful across fields due to its fundamental nature, encapsulated in a quotation from Bernhard Riemann "\textit{In effect, if one extends these functions by allowing complex values for the arguments, then there arises a harmony and regularity which without it would remain hidden.}"~\cite{stein2010complex} For instance, (i) the analysis of the complexified dynamics yielding the Mandelbrot fractals~\cite{branner1989mandelbrot,mandelbrot2004fractals,Lei1990} has pushed the theory of fractals in general, (ii) the analytic continuation of partition functions in Statistical Physics to complex parameters (such as complex temperatures and fields) has led to an analytic theory of phase transitions (see Refs.~\onlinecite{yang_lee1,yang_lee2} by Yang and Lee), and (iii) the complexification of Hamiltonian operators initiated the research field of $\mathcal{PT}$-symmetric quantum mechanics~\cite{Bender_2005,Bender_2007,Bender_2008,Bender_2011,Bender_2021}. Moreover, a recent article suggested to more closely focus on networks with complex weights, also mentioning dynamical systems models and the Kuramoto model~\cite{bottcher2023complex}, yet did not present an analysis of the consequences. Although much of the insights relied on analytically continuing also the system parameters, research on the analytically continued Kuramoto models with complex parameters is missing so far.

In this article, we generalize the original real-variable Kuramoto model and analytically continue it to both complex state variables\cite{Moritz_thuemler2023} and complex parameters. Our findings for $N=2$ coupled units include conditions underlying attracting and non-attracting synchronous states, the coexistence of librating and rotating motions and motion with  finite, yet arbitrarily large rotation numbers, phenomena akin to the undamped and damped physical pendulum~\cite{Strogatz2000Book}. Numerical analysis for networks of $N=128$ also suggests a novel form of discontinuous phase transition at low coupling magnitudes.

These results not only uncover conditions for non-synchronization and synchronization for complexified Kuramoto models that may help better understand finite-$N$ real systems, they also conceptually expand the perspective on the nonlinear dynamics of coupled dynamical systems to their analytic continuation.

\section{\label{sec:complexified_system_SLEE}Kuramoto Model: From Real to Complex} 

\subsection{\label{subsec:real_valued_KM_SLEE}The original, real-variable system}

Let us first consider a system of (real-valued) Kuramoto oscillators that characterizes the collective dynamics of a broad class of coupled phase oscillators~\cite{Acebron2005,RODRIGUES20161}. Such an oscillator is fully described by its phase variable $x_n$ that is either defined as a real number or as an angular coordinate of the 1-torus, i.e. $x_n \in \mathbb{T} := \mathbb{R}/2\pi\mathbb{Z}$ on the unit circle. Their dynamics is governed by
\begin{align}
    \frac{\mathrm{d} }{\mathrm{d} t}x_n = \omega_n + \frac{K}{N} \sum_{m=1}^N \sin(x_m - x_n) 
    \label{eq:KM_real_SLEE} 
\end{align} for $n \in [N] := \{1,2,\cdots,N\}$. In Eq.~(\ref{eq:KM_real_SLEE}), the intrinsic local dynamics of each oscillator is determined by its natural (intrinsic) frequency $\omega_n \in \mathbb{R}$, often drawn from a given probability distribution. The broadness of the natural frequency distribution characterizes the degree of mismatch or heterogeneity of the oscillators. Also, the Kuramoto oscillators are coupled via a sinusoidal function of phase difference with a real coupling strength $K \in \mathbb{R}$. For a positive $K>0$, the coupling between oscillators is attractive, tending to adjust their states and frequencies. For negative $K<0$, the coupling is repulsive such that oscillators tend to be repelled from each other. The collective dynamics of the system of coupled Kuramoto oscillators emerges from an interplay between the heterogeneity in the oscillators' individual intrinsic frequencies $\omega_n$ and the coupling strength., cf. Sec.~\ref{subsec:real_two_oscillator}.

\begin{figure}[t!]
\includegraphics[width=1.0\linewidth]{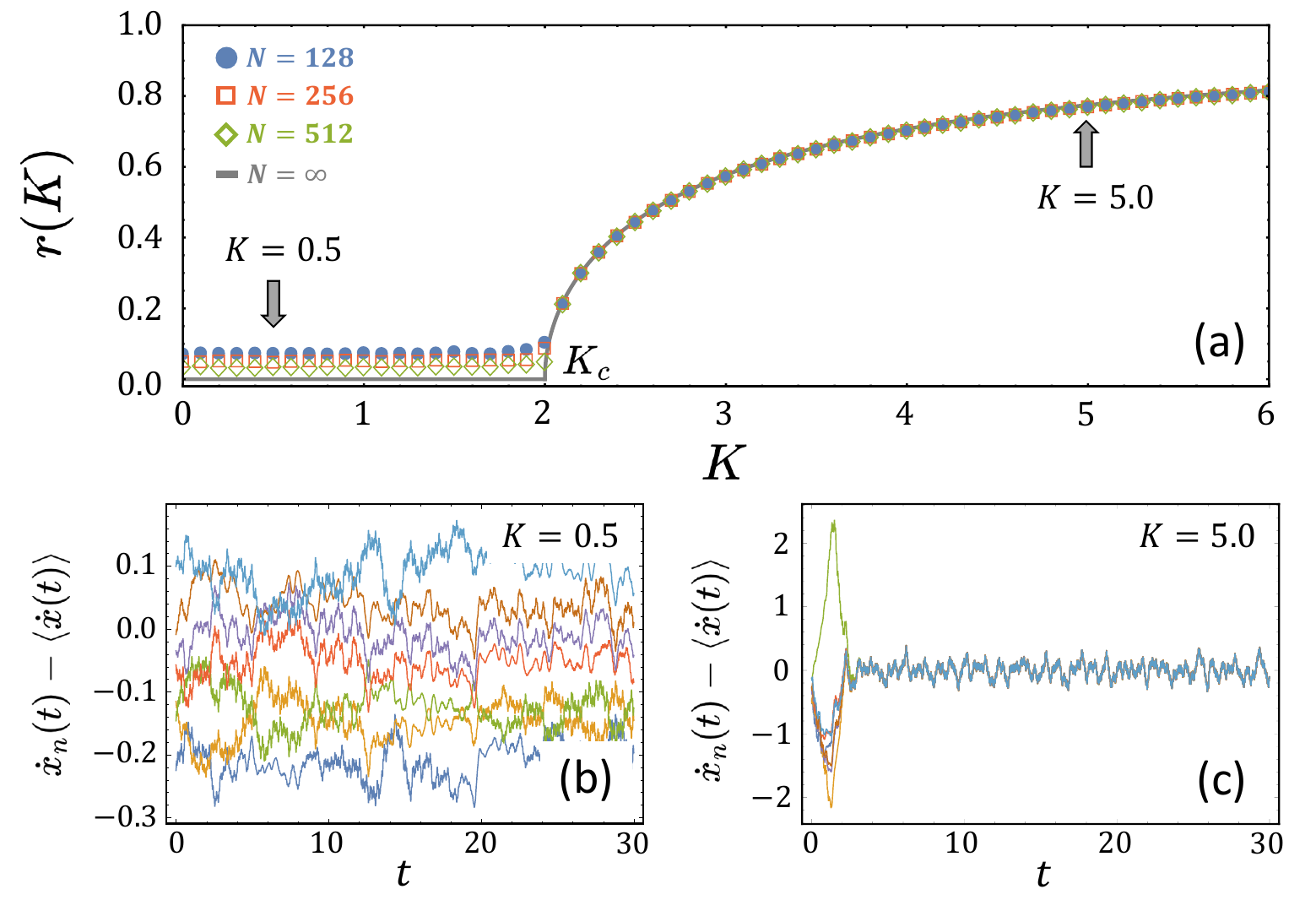}
\caption{ \textbf{Collective dynamics of real-valued Kuramoto oscillators.} (a) The magnitude of the Kuramoto order parameter as a function of the coupling strength $K$ for $N=128$ (blue solid dots), $N=256$ (red open squares), $N=512$ (green open rhombuses), and in the thermodynamic limit (gray solid curve). (b-c) Time evolution of instantaneous velocities of seven selected (of $N=128$) oscillators whose intrinsic frequencies $\omega_n$ are near zero, for (b) $K=0.5 <K_c$ and (c) $K=5.0 >K_c$. Here, the mean instantaneous velocity is defined as $\langle \dot{x}(t) \rangle := \frac{1}{N}\sum_{m=1}^{N}\dot{x}_m(t)$.} 
\label{Fig:real_KM_SLEE}
\end{figure}

As a key observable,  we study the degree of synchrony through the Kuramoto order parameter~\cite{Kuramoto1975,kuramoto2003chemical} 
\begin{align}
    \Gamma(t) := r(t) e^{\textrm{i} \Theta(t)} = \frac{1}{N} \sum_{n=1}^N e^{\textrm{i} x_n(t)} 
    \label{eq:Kuramoto_order_SLEE}
\end{align} with $\textrm{i}:=\sqrt{-1}$  the imaginary unit. The modulus $r(t)$ of $\Gamma(t)$ quantifies the degree of synchrony of oscillators while $\arg(\Gamma(t)):=\Theta(t)$ is the mean phase of the system. For example, $r(t)=1$ marks complete synchronization of all units, i.e., $x_1(t) = x_2(t) = \cdots = x_N(t)$, and small values of $r(t)$ indicate broadly spread state variables $x_n(t)$ at time $t$.

As Figure~\ref{Fig:real_KM_SLEE} illustrates, the real-variable system exhibits a transition from an asynchronous to a synchronous state at some coupling strength $K_c$. Figure~\ref{Fig:real_KM_SLEE} (a) displays the collective behavior of real-valued Kuramoto oscillators in terms of the Kuramoto order parameter $r(t)$ as a function of the coupling strength. Here, a natural frequency is assigned to each oscillator from $\omega_n = \sigma \tan \big(  \frac{\pi(2 n -1-N)}{2N} \big)$ for $n \in [N]$ where $\sigma$ is the scale parameter of a Cauchy-Lorentz distribution. Increasing the coupling strength $K$ from zero to the limit $K \rightarrow \infty$, there is a phase transition in the collective dynamics from incoherence with drifting oscillators and $\left< r(t) \right>_t$ near zero to a partially locked state characterized by $\left< r(t)\right>_t>0$ and eventually to $\left< r(t)\right>_t \rightarrow 1$ in the thermodynamic limit $N\rightarrow \infty$. Here, a time-averaged order parameter is defined as $ \langle r(t) \rangle_t:= \lim_{T \rightarrow \infty} \frac{1}{T} \int_0^T r(t)  \mathrm{d}t$. The Ott-Antonsen ansatz~\cite{OA1,OA2,pikovsky_WS2}, a self-consistency equation~\cite{Strogatz2000,daFonseca_2018}, or Fokker-Planck method~\cite{Strogatz1991} analytically demonstrates a qualitative change in ordering if the coupling strength crosses $K_c = 2\sigma$. Specifically, the Kuramoto order parameter in the thermodynamic limit $N\rightarrow \infty$ (gray solid curve) is $r(K) = 0$, indicating the absence of synchrony for $K<K_c$, and $r(K) = \sqrt{1-\frac{2\sigma}{K}}$ for $K>K_c$ indicating an increasing degree of synchrony in the system.

Different degrees of synchrony also emerge in systems consisting of a finite number $N\in\mathbb{N}$ of units, see Fig.~\ref{Fig:real_KM_SLEE}. For sufficiently strong coupling, phase-locking emerges as a strong form of synchrony where all pairs of units exhibit frequency differences that are zero in time, $\dot{x}_n(t) - \dot{x}_m(t) = 0$. Decreasing $K$ reduces the alignment among the phases, decreasing the order parameter up to a point below which the phase-locked state no longer exists. If the coupling strength decreases further, long-time averages of the order parameter $ \langle r \rangle_t $ gradually decrease to $\mathcal{O}(N^{-1/2})$ as $K\rightarrow 0$. Most of the insights about finite systems, however, are based on direct numerical simulations of the dynamics while several fundamental questions remain open~\cite{Strogatz2000,Acebron2005,RODRIGUES20161,DORFLER20141539} and even posing suitable questions remains a challenge for finite-$N$ systems~\cite{Peter2018,Peter2019,Nishikawa2014_finite}.

\subsection{Complexified Synchrony in the Kuramoto Model}

Recent work has taken a novel perspective and extended the real state variables to become complex by analytic continuation~\cite{Moritz_thuemler2023}. In this article, we moreover complexify the system parameters. Each complexified oscillator $z_n \in \mathbb{C}$ for $n \in [N] $ is governed by
\begin{flalign}
    \frac{\diff }{\diff t} z_n = \omega_n + \frac{K}{N}\sum_{m=1}^{N}\sin(z_m-z_n) \label{eq:KM_complexifed_SLEE}
\end{flalign} where complex velocities $\omega_n \in \mathbb{C}$, generalizations of the intrinsic or natural frequencies, determine the variables' intrinsic dynamics. We write $z_n=x_n+\textrm{i}y_n$, thus introducing $x_n := \Re[ z_n] \in \mathbb{R}$ and $y_n := \Im [z_n] \in \mathbb{R}$ for all $n \in [N]$, and parametrize the complexified coupling constant as
\begin{flalign}
    K = |K|e^{\textrm{i}\alpha} \in \mathbb{C}.
\end{flalign} 
Here $\alpha := \arg (K) \in \mathbb{R}$ takes a role of the phase-lag parameter between real parts of the complex oscillators (see Sec.~\ref{sec:finite_ensemble_SLEE}). The state of such a complexified oscillator can be alternatively defined on an infinite cylinder\cite{Seung_Yeal2012,Seung_Yeal2021} with $(x_n, y_n) \in \mathbb{T} \times \mathbb{R}$  or on $\mathbb{C} \cong \mathbb{R}^2$ with $2\pi$-periodic real parts.

Confining parameters $\omega_n$ and $K$ to be real, the state space of the original, real-valued Kuramoto model constitutes an invariant manifold 
\begin{equation}
\mathrm{M}_0:=\{\mathbf{z} \in \mathbb{C}^N \ | \forall n\in [N] : \ \text{Im}(z_n) =0 \}   \label{eq:invariant_KM_manifold_SLEE} 
\end{equation}
embedded into the full complex state space of Eq.~\eqref{eq:KM_complexifed_SLEE}. As a consequence, all variables stay real, $y_n(t) = 0$  for all times if all initial conditions are from that manifold, $y_n(0)=0$. 
Analogously, for each $y\in \mathbb{R}$, there is an invariant manifold $\textrm{M}_y$ with all imaginary parts the same, $y_n=y$ for all $n\in [N]$.
In the complex domain $\mathbb{C}^N$, the model exhibits fixed points that move off the real manifold if the coupling strength $K$ decreases through $K^\mathrm{(pl)}$ from above\cite{Moritz_thuemler2023}. 
We emphasize that such complex fixed points still exist for $K<K^\mathrm{(pl)}$; in contrast, phase-locked states represented by real fixed points disappear in the real-variable model. These \textit{complex locked states} constitute a new form of \textit{complexified synchrony} that moreover, is analytically accessible. In particular, the original work in Ref.~\onlinecite{Moritz_thuemler2023} suggests that stability of the (completely analytically determined) complex locked states implies phase-locking in the original, real-variable model.

\section{\label{sec:Two_model_SLEE}A System of Two Complexified Oscillators}

\subsection{\label{subsec:real_two_oscillator}Two Real-valued Kuramoto Oscillators: Revisited}

Let us first briefly revisit two coupled original (real-variable) Kuramoto oscillators. The system (\ref{eq:KM_real_SLEE}) has a phase-shift invariance since oscillators are coupled through their phase differences. Thus, in Eq.~(\ref{eq:KM_real_SLEE}) with $N=2$, the phase difference $\Delta x := x_2-x_1$ follows
\begin{flalign}
    \frac{\diff}{\diff t}\Delta x = \Delta \omega - K \sin\Delta x =: f(\Delta x) \label{eq:real_two_SLEE}
\end{flalign} 
where $\Delta \omega := \omega_2-\omega_1 \in \mathbb{R}$ and $K \in \mathbb{R}$. In Eq.~(\ref{eq:real_two_SLEE}), $\Delta x =0$ is not a fixed point unless $\Delta\omega=0$. As in general $f(0) \neq 0$, complete synchrony of two oscillators ($x_1(t)=x_2(t)$ for all $t>0$) does not constitute an invariant state. Complete synchronization occurs only if either the oscillators have identical frequencies, i.e., $\Delta \omega=0$, or in the limit of infinitely strong coupling, $K \rightarrow \infty$. Nevertheless,  frequency synchronization and thereby phase-locking emerges for sufficiently large $K$, such that two oscillators are entrained at the same frequency $\dot{x_1}=\dot{x_2}$, and $\frac{\diff}{\diff t}\Delta x = f(\Delta x)=0$. We remark that two locked states emerge as fixed points to Eq.~(\ref{eq:real_two_SLEE}): $\Delta x_0^* = \sin^{-1}\frac{\Delta \omega}{K}$ or $\Delta x_1^* = \pi- \sin^{-1}\frac{\Delta \omega}{K}$. The former constitutes an in-phase state where the phase difference decreases and converges to $\Delta x_0^* \rightarrow 0$ as $K\rightarrow\infty$ whereas the latter represents an anti-phase state with the two phases nearly opposite to each other on the unit circle and converging to $\Delta x^*_1 \rightarrow \pi$ as $K\rightarrow\infty$. Both phase-locked states exist if and only if $|\frac{\Delta\omega}{K}| \leq 1$, meaning that the system is required to have either sufficiently strong coupling for a given mismatch in intrinsic frequencies or a small frequency mismatch at a given coupling strength. In contrast, no fixed points exist for $|\frac{\Delta\omega}{K}| > 1$, and thus the oscillators are drifting relative to each other, i.e., the system shows an incoherent state. For $|\frac{\Delta\omega}{K} |\leq 1$, a locked state is linearly stable if
\begin{equation}
    \frac{\diff f}{\diff \Delta x}\bigg|_{\Delta x = \Delta x_{0,1}^*} = -K \cos\Delta x_{0,1}^* = \mp K \sqrt{1-\frac{\Delta\omega^2}{K^2}}<0. 
\end{equation} For $K >0$, i.e., an attractive coupling, the locked state $\Delta x_0^*$ representing nearly in-phase synchrony is stable (and unstable for repulsive coupling $K<0$). Thus, positive (attractive) coupling induces `attraction' of the oscillators towards each other and thus towards a common collective phase whereas negative (repulsive) coupling induces repulsion of the oscillators away from each other, such that $\Delta x_1^*$ is stable and $\Delta x_0^*$ is unstable. For more details about the two-oscillator systems, whose differential equation for the phase difference is also known as the Adler equation~\cite{Pikovsky2007_scholar}, see Sec. 8.6 in Ref.~\onlinecite{Strogatz2000Book}.

\subsection{\label{subsec:complex_two_oscillator}Two Complexified Kuramoto Oscillators}

Let us now study a system of two complexified oscillators, i.e., $N=2$ in Eq.~(\ref{eq:KM_complexifed_SLEE}). The anti-symmetric state-difference coupling reduces \eqref{eq:KM_complexifed_SLEE}  to one ordinary differential equation for the difference $\Delta z := z_2-z_1$  of the two complex variables. Taking the parameters $K=|K|e^{\textrm{i}\alpha}$ and defining  $\Delta \omega := \omega_2 - \omega_1 = |\Delta \omega |e^{\textrm{i}\gamma} \in \mathbb{C}$ with $\gamma := \arg(\Delta\omega) \in \mathbb{R}$, the dynamics follows
\begin{flalign}
    \frac{\diff}{\diff t} \Delta z &= \Delta \omega - K \sin \Delta z \notag \\
    &= |\Delta \omega|e^{\textrm{i}\gamma} - |K|e^{\textrm{i}\alpha}\sin\Delta z 
\label{eq:two_oscillator_general_parameter_SLEE}
\end{flalign} where $|K|$ determines how strong the coupling between two oscillators is, while $|\Delta \omega|$ characterizes the degree of heterogeneity of the system. To unify the analysis, we hereafter explore two-oscillator systems with time rescaled by the  coupling strength $|K|$ , with their dynamics governed by
\begin{flalign}
   \frac{\diff}{\diff t} \Delta z &= c e^{\textrm{i}\gamma} - e^{\textrm{i}\alpha}\sin\Delta z . \label{eq:two_general_rescaledtime_SLEE}
\end{flalign} 
Here 
\begin{equation}
    c = \frac{|\Delta \omega|}{|K|} \geq 0
\end{equation}
is a bifurcation parameter that characterizes the inhomogeneity $|\Delta \omega|$ of the oscillators' intrinsic dynamics relative to the strength $|K|$ of the coupling between them. 

The dynamics \eqref{eq:two_general_rescaledtime_SLEE} is controlled by three (scalar, real) effective parameters:  $c$, $\gamma$, and $\alpha$. In the following, we analyze the influence of each parameter on the system of two complexified oscillators.

\subsection{\label{subsec:real_omega_SLEE}Real Natural Frequencies and Complex Coupling}

First, consider a system of two complexified oscillators with real natural frequencies ($\omega_1,\omega_2 \in \mathbb{R}$ and so $\gamma=0$ in Eq.~\eqref{eq:two_general_rescaledtime_SLEE}). The governing equation then reads
\begin{flalign}
    \frac{\diff }{\diff t} \Delta z = c - e^{\textrm{i}\alpha}\sin\Delta z =: f(\Delta z)\label{eq:two_real_natural_SLEE}
\end{flalign} where $\alpha \in \mathbb{R}$. Via properties of a complex sine function, the governing equation (\ref{eq:two_real_natural_SLEE}) becomes
\begin{flalign}
    \frac{\diff}{\diff t}\Delta x &=c + \sin\alpha \cos \Delta x \sinh \Delta y -\cos\alpha \sin\Delta x \cosh \Delta y \notag \\
    \frac{\diff }{\diff t}\Delta y &= -\cos\alpha \cos \Delta x \sinh \Delta y -\sin\alpha \sin\Delta x \cosh \Delta y \label{eq:rescaled_two_difference_SLEE}
\end{flalign} 
in terms of the real and imaginary parts of the state variable $\Delta z=\Delta x + \textrm{i} \Delta y$. Depending on the value of $\alpha$,  the system (\ref{eq:two_real_natural_SLEE}) exhibits three qualitatively different dynamics: (i) for real-valued coupling for $\alpha=0$ complex locked states emerge replacing real locked states below some critical coupling $K^{\text{(pl)}}$, as uncovered in Ref.~\onlinecite{Moritz_thuemler2023}. In the following subsections, we discuss systems with (ii) purely imaginary coupling where $\alpha = \frac{\pi}{2}$ and (iii) generic complex coupling where $\alpha \in (0, \frac{\pi}{2})$.

\subsubsection{\label{subsec:Two_model_purely_SLEE}Purely Imaginary Coupling: $\alpha = \frac{\pi}{2}$}

\begin{figure*}[t!]
\includegraphics[width=0.63\textwidth]{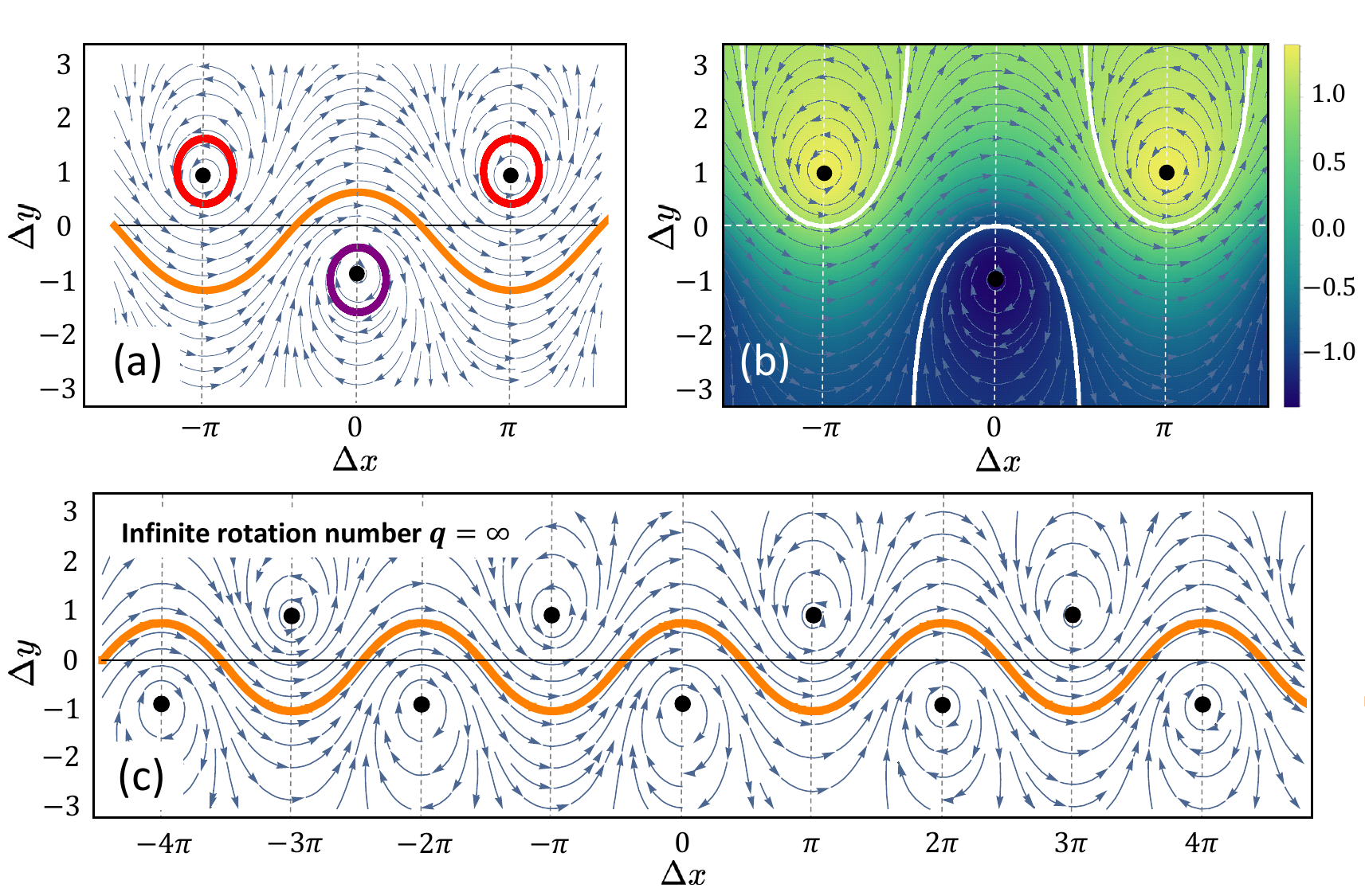}
\caption{ \textbf{Libration and rotation for purely imaginary coupling ($N=2$).} (a) A phase portrait around three fixed points. Black solid dots indicate fixed points (\ref{eq:fixedpoints_pure_im_SLEE}). The orange solid curve indicates a \textit{rotation} (with the rotation number $q=\infty$) whereas the red curves display \textit{librations}  ($q=0$) encircling antiphase fixed points and the purple curve indicates a libration around an in-phase synchrony fixed point. (b) The same fixed points as in (a) with separatrices (white solid curves) and the conserved quantity defined in Eq.~(\ref{eq:conserved_im_SLEE}). (c) A trajectory that rotates along the cylindrical axis with an infinite rotation number $q=\infty$. All subfigures are obtained for $c=1$.} 
\label{Fig:purely_imaginary1_SLEE}
\end{figure*}

For purely imaginary coupling, where $\alpha := \arg(K) = \frac{\pi}{2}$, the governing equations become
\begin{flalign}
     \frac{\diff}{\diff t}\Delta x &=c + \cos \Delta x \sinh \Delta y \notag , \\
    \frac{\diff }{\diff t}\Delta y &=- \sin\Delta x \cosh \Delta y .\label{eq:pure_im_two_difference_SLEE}
\end{flalign} Two fixed-point solutions are given by
\begin{flalign}
  (\Delta x_0^*, \Delta y_0^*) &= (0, -\sinh^{-1}c) \notag \\
  (\Delta x_1^*, \Delta y_1^*) &= (\pi, \sinh^{-1}c)
     \label{eq:fixedpoints_pure_imag_SLEE}
\end{flalign} such that they are $\Delta z_1^* = \pi + \textrm{i}\sinh^{-1}c$ and $\Delta z_0^* = - \textrm{i}\sinh^{-1}c$ in  complex number form. Since the complex vector field (\ref{eq:two_real_natural_SLEE}) satisfies the Cauchy-Riemann condition, the two eigenvalues of the Jacobian matrix of \eqref{eq:pure_im_two_difference_SLEE} are obtained as
\begin{flalign}
    \lambda_+ &= \frac{\partial}{\partial \Delta x} \Re[f] + \textrm{i}\frac{\partial}{\partial \Delta y} \Re[f] = f'(\Delta z) \notag \\
    \lambda_- &= \frac{\partial}{\partial \Delta x} \Re[f] - \textrm{i}\frac{\partial}{\partial \Delta y} \Re[f] = \overline{f'(\Delta z)} \label{eq:linear_stability_SLEE}
\end{flalign} where an overbar indicates complex conjugation. For $\alpha=\frac{\pi}{2}$, the eigenvalues of the Jacobian matrix evaluated at a fixed point $\Delta z^*$ are given by $f'(\Delta z^*) = -\textrm{i}\cos\Delta z^*$, which yields $f'(\Delta z_1^*) = \textrm{i}\sqrt{1+c^2}$ and $f'(\Delta z_0^*) = -\textrm{i}\sqrt{1+c^2}$ together with their complex conjugates. Hence, both the two fixed points in Eq.~(\ref{eq:fixedpoints_pure_imag_SLEE}) are linearly neutrally stable for any $c \in \mathbb{R}$. Moreover, since the sinusoidal function, $\sin \Delta x$ is  $2\pi$-periodic in $\Delta x$, the system (\ref{eq:pure_im_two_difference_SLEE}) has infinitely many fixed point solutions in the complex plane,
\begin{flalign}
    (\Delta x_{2k}^*, \Delta y_{2k}^*) &= (0 +2 \pi k, -\sinh^{-1}c) \notag \\
    (\Delta x_{2k+1}^*, \Delta y_{2k+1}^*) &= (\pi + 2 \pi k, \sinh^{-1}c) \label{eq:fixedpoints_pure_im_SLEE}
\end{flalign} for all $k \in \mathbb{Z}$.

Thus, both eigenvalues \eqref{eq:linear_stability_SLEE} are purely imaginary for all fixed points so that all of them are neutrally stable (centers). We emphasize that all the fixed points $\Delta z^*_{2k}$ with even indices represent complete in-phase synchrony with respect to their real parts, i.e., $\Delta x_{2k}=0$ (mod $2\pi$) whereas those with odd indices, $\Delta z^*_{2k+1}$, represent exact antiphase or splay states with  $\Delta x_{2k+1}=\pi$ (mod $2\pi$). Thus, in contrast to the original, real-valued model, the complexified system possesses (neutrally stable) states exhibiting \textit{complete in-phase synchrony even though the intrinsic frequencies are different}, $\Delta\omega\neq0$ and the coupling strength is finite, $|K| <\infty$.

Moreover, numerical integration of Eq.~(\ref{eq:pure_im_two_difference_SLEE}) indicates that the system exhibits two and only two different types of periodic motion if initialized from random initial conditions, see Fig.~\ref{Fig:purely_imaginary1_SLEE} (a). First, an infinite number of non-isolated periodic orbits exist around each fixed point in Eq.~(\ref{eq:fixedpoints_pure_im_SLEE}). Considering phase space as the surface of a cylinder, such periodic motions constitute \textit{librations} that encircle one given fixed point. Second, a topologically different type of periodic motion exists further away from the fixed points, exhibiting indefinite \textit{rotation} on the cylindrical surface around the longitudinal $\Delta x$-axis. 

\begin{figure*}[t!]
\includegraphics[width=0.63\textwidth]{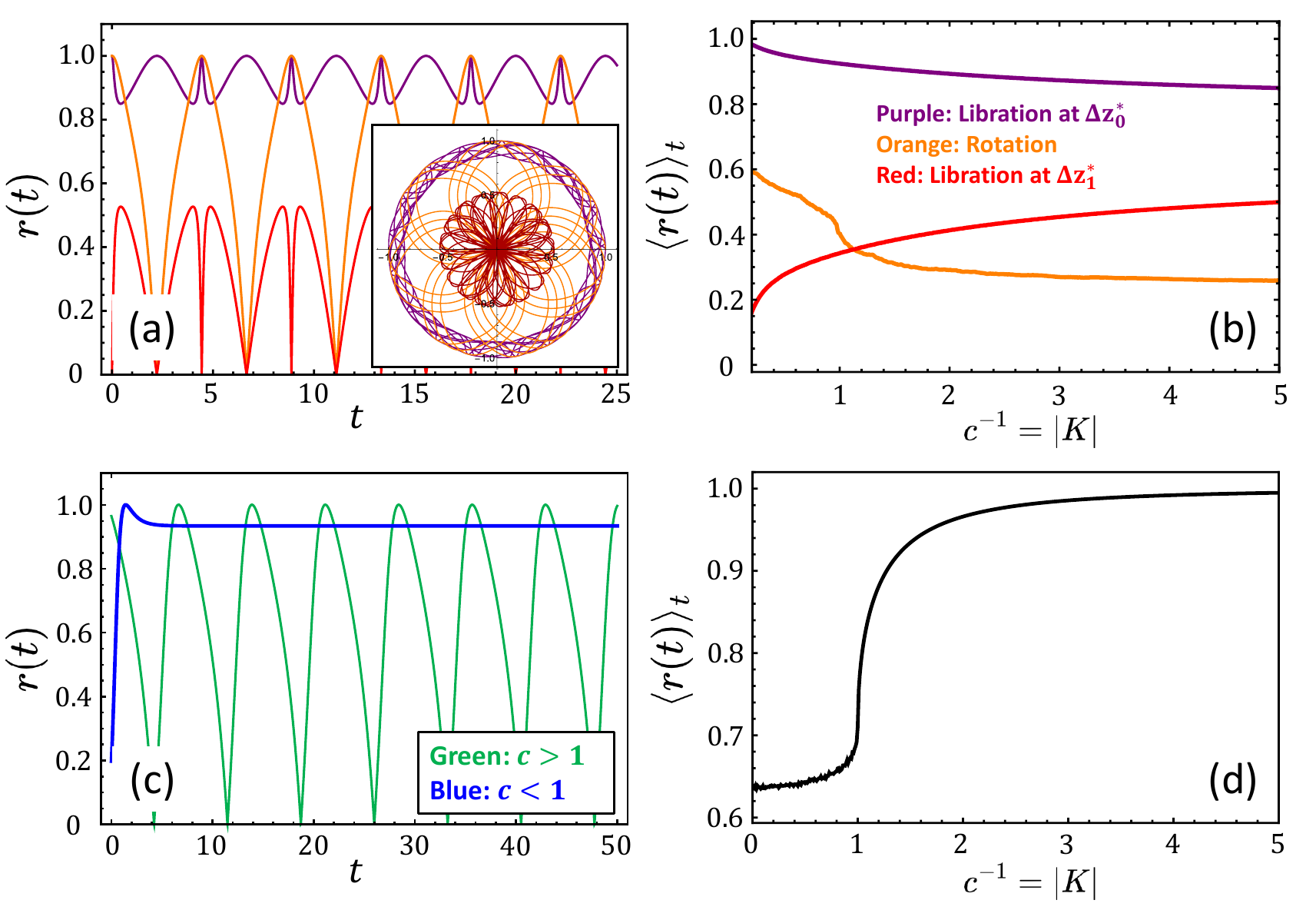}
\caption{\textbf{Absence of transition to locking ($N=2$)}.
Comparison between the Kuramoto model with panels a),b), purely imaginary coupling ($\alpha=\frac{\pi}{2}$) and c),d), the real-valued Kuramoto model ($\alpha=0$).  (a) Time evolution of the magnitude $r(t)$ of the complex Kuramoto order parameter for a rotation (orange) and a libration around a locked state: in-phase synchrony $\Delta z^*_0$ (purple) and an antiphase state $\Delta z^*_1$ (red). Here, $c=1$. Inset: the complex Kuramoto order parameter (\ref{eq:Kuramoto_order_SLEE}) in the complex plane. (b) Time-averaged $\left<r(t)\right>_t$ as a function of the coupling strength $|K|$. For every $c$, each data point is obtained from an initial condition chosen close to the respective type of solution (libration around one or another fixed point or rotation). (c) Time evolution $r(t)$ for $c>1$ (no locked state) and $c<1$ (locked state). (d) Time-averaged $\left<r(t)\right>_t$ as a function of the coupling strength $K>0$ for $c<1$. Data points are obtained from  initial conditions randomly drawn from $\mathbb{T}^2$. For all panels, we set $\Delta \omega=1$ and vary $c$ via $|K|$. }
\label{Fig:purely_imaginary2_SLEE}
\end{figure*}

As the numerical results and the linear stability analysis suggest, the neutrally stable fixed points arise due to the existence of a conserved quantity. Indeed, an energy function defined as
\begin{flalign}
    E(\Delta x, \Delta y) := \frac{-\cos \Delta x}{\cosh \Delta y} + c \tanh \Delta y \label{eq:conserved_im_SLEE}
\end{flalign} is conserved, 
\begin{equation}
    \frac{\diff}{\diff t}  E(\Delta x, \Delta y) =0,
\end{equation} along any trajectory of the system (\ref{eq:pure_im_two_difference_SLEE}), irrespective of the parameter 
$c = \frac{|\Delta \omega|}{|K|}$, in particular also for both librating and rotating motion. At all fixed points, the gradient of the energy function is the zero vector, 
\begin{equation}
    \nabla E(\Delta x^*, \Delta y^*) = \left(\frac{\sin\Delta x^*}{\cosh\Delta y^*}, \frac{c+\cos\Delta x^* \sinh\Delta y^*}{\cosh^2 \Delta y^*}\right)^\top = (0,0)^\top.
\end{equation} 
Moreover, its Hessian matrix
\begin{equation}
    \bold{H}E(\Delta x_{2k+1}^*, \Delta y_{2k+1}^*) = -\frac{1}{\sqrt{1+c^2}}I_2
\end{equation} with identity matrix $I_2 \in \mathbb{R}^{2\times 2}$ is negative-definite while 
\begin{equation}
    \bold{H}E(\Delta x_{2k}^*, \Delta y_{2k}^*) = \frac{1}{\sqrt{1+c^2}}I_2
\end{equation} is positive-definite. Thus, the fixed points of in-phase synchrony $(\Delta x_{2k}^*, \Delta y_{2k}^*)$ are local minima of $E(\Delta x, \Delta y)$ whereas fixed points of antiphase states $(\Delta x_{2k+1}^*, \Delta y_{2k+1}^*)$ are local maxima.
As Fig.~\ref{Fig:purely_imaginary1_SLEE} (b) illustrates, separatrices in state space separate trajectories with librating motion and thus rotation number $q=0$  from those with rotating motion and rotation number $q=\infty$, see also Fig.~\ref{Fig:purely_imaginary1_SLEE} (c). Here, the rotation number is defined as the integer number of turns that the rotational trajectory takes along the $\Delta x$-axis around the cylinder, $\Delta x \in \mathbb{T}$, i.e., the integer multiple
$$
 q = \lim_{t\rightarrow\infty} \bigg\lfloor \frac{\Delta x(t) - \Delta x(0)}{2 \pi}\bigg\rfloor
$$
of $2 \pi$ that the variable $\Delta x$ increases along the trajectory. Here $\lfloor . \rfloor$ is the floor function.

Thus, for $N=2$  coupled Kuramoto oscillators with a purely imaginary coupling, no classical synchronization occurs, even for arbitrarily large coupling strength $|K|\rightarrow \infty$ and arbitrarily small differences in the intrinsic frequencies $\Delta\omega >0$.
In particular, trajectories starting from almost all initial conditions
do not reach any of the fixed points, $\{(\Delta x^*_k,\Delta y^*_k) | k\in \mathbb{Z}\} \subset \mathbb{T}\times \mathbb{R}$, because they are neutrally stable and thus non-attracting. Figure~\ref{Fig:purely_imaginary2_SLEE} illustrates that the order parameter as a function of $c$ (and thus $|K|$) exhibits no transition for the complexified model, in contrast to the original real-variable model.

\subsubsection{\label{subsec:Two_model_complex_SLEE}Generic Complex Coupling: $\alpha \in (0, \frac{\pi}{2})$}

We now study the dynamics of systems with  generic complex coupling,  $K=|K|e^{\textrm{i}\alpha} \in \mathbb{C}$ for $\alpha \in (0, \frac{\pi}{2})$. 
\vspace*{1mm}
\begin{figure*}[t!]
\includegraphics[width=0.63\textwidth]{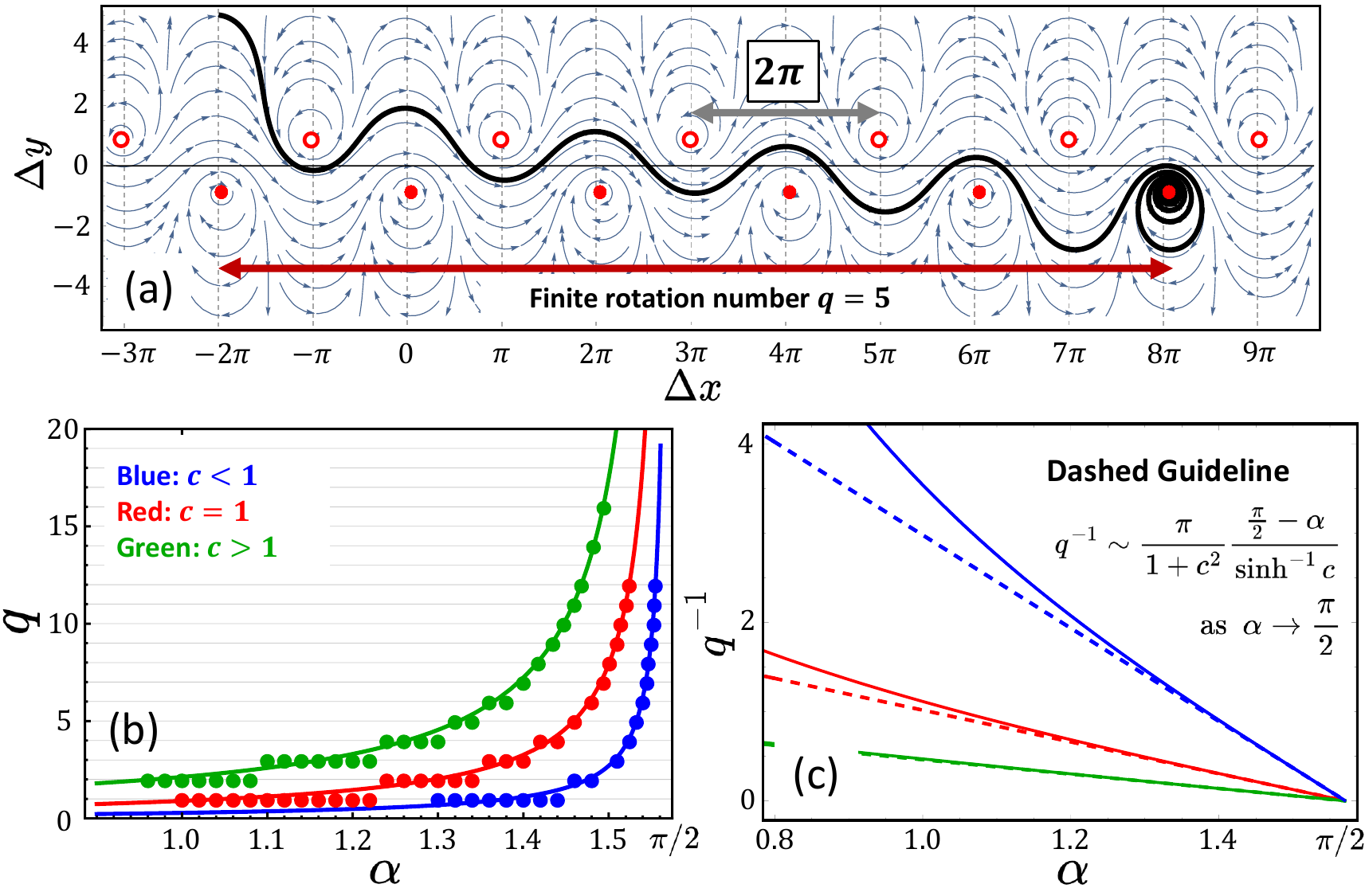}
\caption{\textbf{Full range of integer rotation numbers $q\in\mathbb{N}$ for generic  complex coupling ($N=2$)}. (a) A trajectory (black solid curve) for $c=1$ and $\alpha = 1.46004$ is depicted that circles around the cylinder with a finite rotation number $q=5$ and eventually stops at a stable fixed point. Red dots indicate fixed points: stable (solid) in Eq.~(\ref{eq:stable_fixed_comp_SLEE}) and unstable (open) in Eq.~(\ref{eq:Unstable_fixed_comp_SLEE}). (b) The analytical form (solid curves) of the rotation number $q$ is depicted as a function of the parameter $\alpha$, together with numerical results (solid dots). (c) The asymptotic behavior (dashed line, Eq.~(\ref{eq:asymp_winding_SLEE})) of the rotation number $q$ as $\alpha \rightarrow \frac{\pi}{2}^-$: blue ($c=0.5<1$), red ($c=1$) and green ($c=1.5>1$).} 
\label{Fig:complex_SLEE}
\end{figure*}

The two-unit system \eqref{eq:rescaled_two_difference_SLEE} with $\alpha \in (0, \frac{\pi}{2})$ possesses fixed points
\begin{flalign}
    \Delta x_{2k}^* &= 2\pi k+\sin^{-1}\Bigg(\sqrt{\frac{1+c^2-\sqrt{1+c^4-2c^2\cos2\alpha}}{2}} \Bigg)  \notag \\
    \Delta y_{2k}^* &= -\sinh^{-1} \Bigg( \frac{\sqrt{2}c\sin\alpha}{\sqrt{1-c^2+\sqrt{1+c^4-2c^2\cos2\alpha}}}\Bigg) <0 \label{eq:stable_fixed_comp_SLEE}
\end{flalign} 
for $k\in\mathbb{Z}$
that represent near in-phase states in their real parts ($\Delta x^*_{2k}=0$), in particular $\Delta x^*_{2k} = c \cos\alpha +\mathcal{O}(c^3)$ (mod $2 \pi$) asymptotically as $c \rightarrow 0$. In addition, the other fixed points, representing (near-) anti-phase states, are given as
\begin{flalign}
    \Delta x_{2k+1}^* &= \pi-\Delta x _{2k}^* \notag \\
    \Delta y _{2k+1} ^* &= -\Delta y_{2k}^* >0\label{eq:Unstable_fixed_comp_SLEE}
\end{flalign} for $k\in\mathbb{Z}$. Linear stability analysis from \eqref{eq:linear_stability_SLEE} shows that the complex locked states \eqref{eq:stable_fixed_comp_SLEE} are linearly stable whereas the states \eqref{eq:Unstable_fixed_comp_SLEE} are unstable. A bifurcation takes place at $\alpha = \frac{\pi}{2}$ wherein neutrally stable fixed points associated with in-phase synchrony are stabilized. Indeed, at a given fixed point $\Delta z^*$, the Jacobian has eigenvalues $f'(\Delta z^*) = -e^{\textrm{i}\alpha}\cos \big(\sin^{-1}(c e^{-\textrm{i}\alpha}) \big)$ for $\Delta z^* = \sin^{-1}(c e^{-\textrm{i}\alpha})$. Their real parts become zero if $\Im[e^{2\textrm{i}\alpha}-c^2 ] = \sin2\alpha =0$, i.e. for $\alpha =0$ or $\frac{\pi}{2}$ and integer multiples thereof. For more details, see below Eq.~(\ref{eq:stablilt_aympt_SLEE}).

For a complex coupling with $\alpha \in (0, \frac{\pi}{2}) $, the system possesses stable locked states (\ref{eq:stable_fixed_comp_SLEE}). Moreover, a trajectory from a random initial condition does not exhibit a ceaseless rotation nor a libration around a fixed point, but rather it eventually reaches one of the stable equilibria after rotating around the cylindrical axis, i.e., with a finite rotation number $q < \infty$. To see this, consider a trajectory starting from an initial condition that reads $\Delta x(0) = \Delta x_{2k}^*$ and $\Delta y(0) \gg 1$ for a given $k$. In Fig.~\ref{Fig:complex_SLEE} (a), a trajectory initiated from $\Delta x(0) = \Delta x_{-2}^*$ and $\Delta y(0) \gg 1$ is depicted (black solid curve) for $c=1$ and $\alpha < \frac{\pi}{2}$. The trajectory winds around the cylinder and eventually spirals down to a stable locked state, i.e., $(\Delta x_{4}^*, \Delta y_{4}^*)$. Thus, the specific trajectory takes the finite rotation number $q=5$. We emphasize that the rotation number of such a trajectory decreases down to  (any)  finite integer $q<\infty$ from $q=\infty$ as the phase-lag parameter decreases from $\alpha=\frac{\pi}{2}$ (purely imaginary coupling; $K \in \textrm{i}\mathbb{R}$) to $\alpha <\frac{\pi}{2}$ (complex coupling; $K\in \mathbb{C}\setminus \textrm{i}\mathbb{R}$).

Recall that from a numerical integration for $\alpha=\frac{\pi}{2}$, e.g., in Fig.~\ref{Fig:purely_imaginary1_SLEE} (c), a trajectory around the $\Delta x$-axis exhibits a nearly sinusoidal curve with the amplitude $|\Delta y^*| = \sinh^{-1}c$. Also, a numerical integration for $\alpha \lesssim \frac{\pi}{2}$ with the initial condition described above shows a tilted sinusoidal trajectory near the $\Delta x$-axis (see Fig.~\ref{Fig:complex_SLEE} (a)). Thus, we assume the intermediate trajectory near the horizontal axis satisfies $\Delta y = -|\Delta y^*|\cos(\Delta x - \Delta x^*) + A \Delta x= -\sinh^{-1}(c) \cos(\Delta x - \Delta x^*) +A \Delta x$ with a slope $A$ and shifted horizontally by $\Delta x^*$. Also, assuming that the trajectory settles down to a stable fixed point $(\Delta x^*+2\pi k,\Delta y^*)$ without spiraling down to it leads to $\Delta y^* = -\sinh^{-1}c +A \Delta x^*$. Then, the slope $A$ determines the rotation number via $A = \frac{\Delta y^*}{2\pi q} = \frac{\Delta y^* + \sinh^{-1}c}{\Delta x^*}$. Solving this algebraic equation, an analytical form of the rotation number as a function of parameters reads
\begin{widetext}
\begin{flalign}
    q = \frac{1}{2\pi} \frac{ \Delta x^* \Delta y^*}{\Delta y^* + \sinh^{-1}c} 
    =\frac{1}{2\pi} \frac{\sin^{-1}\bigg(\sqrt{\frac{1+c^2-\sqrt{1+c^4-2c^2\cos2\alpha}}{2}} \bigg) \sinh^{-1} \bigg( \frac{\sqrt{2}c\sin\alpha}{\sqrt{1-c^2+\sqrt{1+c^4-2c^2\cos2\alpha}}}\bigg)}{\sinh^{-1}c-\sinh^{-1} \bigg( \frac{\sqrt{2}c\sin\alpha}{\sqrt{1-c^2+\sqrt{1+c^4-2c^2\cos2\alpha}}}\bigg)} \label{eq:winding_number_SLEE}
\end{flalign}     
\end{widetext} for $\alpha < \frac{\pi}{2}$. In Fig.~\ref{Fig:complex_SLEE} (b), the analytic form (\ref{eq:winding_number_SLEE}) of the rotation number is depicted together with numerically obtained results for three different values of $c$. The actual rotation number is an integer and thus a step function of $\alpha$. For the purpose of illustration, we therefore plot it (only) for the phase-lag parameter $\alpha$ that is numerically closest to the analytic curve. The good agreement between the numerical results and analytical prediction means that our analytical estimate \eqref{eq:winding_number_SLEE} is very close to the actual rotation number step function. 
We remark that our prediction \eqref{eq:winding_number_SLEE}
relies on the approximation of the rotating motion by an exact (tilted) sinusoidal function $\Delta y(\Delta x)$.

We underline that the rotation number $q(\alpha)$ in Eq.~(\ref{eq:winding_number_SLEE}) diverges as $\alpha \rightarrow \frac{\pi}{2}^-$, consistent with the findings in the system featuring a purely imaginary coupling, i.e., $\alpha=\frac{\pi}{2}$ where $q=\infty$ in Fig.~\ref{Fig:purely_imaginary1_SLEE} (c). Employing an asymptotic series expansion~\cite{Bender1978} with respect to orders of $\frac{\pi}{2}-\alpha$, the leading order of the asymptotic behavior of the rotation number reads
\begin{flalign}
    q^{-1} \sim \frac{\pi}{1+c^2} \frac{\frac{\pi}{2}-\alpha}{\sinh^{-1}c} \label{eq:asymp_winding_SLEE}
\end{flalign} as $\alpha \rightarrow \frac{\pi}{2}^-$. In Fig.~\ref{Fig:complex_SLEE} (c), we show the comparison of the analytic form of the rotation number (\ref{eq:winding_number_SLEE}) with Eq.~(\ref{eq:asymp_winding_SLEE}) near $\alpha = \frac{\pi}{2}$ from below. For any given $c$, the asymptotic behavior of the rotation number follows Eq.~(\ref{eq:asymp_winding_SLEE}) for $\alpha \to \frac{\pi}{2}^{-}$, i.e., as the complex coupling approaches purely imaginary coupling.

\subsubsection{\label{subsubsec:asymp}Stabilization of Synchrony}

As previously explained, in a system comprising two complexified oscillators with a purely imaginary coupling, there exists no classical locked state from any arbitrary initial condition unless we initialize the system precisely at the fixed points, all of which are neutrally stable. To demonstrate the stabilization of neutrally stable in-phase synchrony and the destabilization of a neutrally stable antiphase state upon the activation of the real part of the complex coupling, we introduce a small perturbation to the phase-lag parameter, i.e., $\alpha=\frac{\pi}{2}-\beta$ for $0<\beta\ll 1$. Consider the governing equation (\ref{eq:two_real_natural_SLEE}) in the first order of $\beta$, which reads
\begin{flalign}
    \frac{\diff}{\diff t} \Delta z &= c - e^{\textrm{i}\alpha}\sin\Delta z \notag \\
    &\sim  c - \textrm{i} (1-\textrm{i}\beta) \sin\Delta z =: f(z) \label{eq:pertur_eq_SLEE}
\end{flalign} as $\beta \rightarrow 0^+$. Exploiting an asymptotic series expansion for a fixed point solution, i.e., $\Delta z^* \sim a_0 + \sum_{n=1}^{\infty}a_n \beta^n$ as $\beta \rightarrow 0^+$ where $0= c -\textrm{i}(1-\textrm{i}\beta)\sin\Delta z^*$, the zeroth order coefficient is a fixed point solution for the purely imaginary case (\ref{eq:fixedpoints_pure_im_SLEE}), i.e., $\sin a_0 = -\textrm{i}c $. Then, the first order coefficient is given as $a_1 = \textrm{i}\tan a_0 = \pm \frac{c}{\sqrt{1+c^2}}$. The equilibrium in the series expansion then reads
\begin{flalign}
 \Delta z^* &= a_0 \pm \frac{c}{\sqrt{1+c^2}}\beta + \mathcal{O}(\beta^2) \notag \\
 &= \left\{ \begin{array}{cl}
-\sin^{-1}(\textrm{i}c)+\frac{c}{\sqrt{1+c^2}}\beta + \mathcal{O}(\beta^2) & : \ \Delta z^*_{2k} \\
\pi+\sin^{-1}(\textrm{i}c)-\frac{c}{\sqrt{1+c^2}}\beta + \mathcal{O}(\beta^2) & : \   \Delta z^*_{2k+1}
\end{array} \right. \label{eq:asymp_fixed_SLEE}   
\end{flalign} for $k \in \mathbb{Z}$ as $\beta \rightarrow 0^+$. The linear stability of the fixed points $\Delta z^*$ in Eq.~(\ref{eq:asymp_fixed_SLEE}) is analyzed using Eq.~(\ref{eq:linear_stability_SLEE}). The eigenvalues of the Jacobian matrix evaluated at the equilibrium are given as
\begin{flalign}
    f'(\Delta z^*) &= -\textrm{i}(1-\textrm{i}\beta) \cos \Delta z^* \notag \\
    &= -\textrm{i}\cos a_0 \pm \textrm{i}\frac{c}{\sqrt{1+c^2}}\beta \sin a_0  -  \beta  \cos a_0 + \mathcal{O}(\beta^2)
    \notag \\
    &= \mp \textrm{i}\sqrt{1+c^2} \mp \frac{1}{\sqrt{1+c^2}}\beta +\mathcal{O}(\beta^2) \label{eq:stablilt_aympt_SLEE}
\end{flalign} as $\beta \rightarrow 0 ^+$. The first term reflects the neutral stability of the zeroth order, i.e., neutrally stable fixed points for a purely imaginary coupling. It is the second term that allows the fixed points to be stabilized for $\Im[\Delta z^*]<0$ and destabilized for $\Im[\Delta z^*]>0$, respectively. Therefore, we deduce that even a slight activation of the real part of the complex coupling stabilizes the equilibria associated with phase synchrony, enabling the observation of classical locked states for $\beta >0$ with a finite coupling strength $|K| \in \mathbb{R}$.

\subsection{\label{subsec:general_two_SLEE}Complex Natural Frequencies and Complex Coupling}

So far we have investigated a system of two complexified oscillators with complex coupling constant $K \in \mathbb{C}$ and real natural frequency $\Delta \omega \in \mathbb{R}$. We now briefly discuss a two-oscillator system where also the  natural frequencies in Eq.~(\ref{eq:two_general_rescaledtime_SLEE}) are complex, i.e., $\Delta \omega = |\Delta \omega|e^{\textrm{i}\gamma} \in \mathbb{C}$ for $\gamma \neq 0$. From Eq.~(\ref{eq:two_general_rescaledtime_SLEE}) we obtain fixed point solutions via
\begin{flalign}
    0 &= c -e^{\textrm{i}(\alpha-\gamma)}\sin\Delta z, \notag \\
   0 &= c -e^{\textrm{i}\hat{\alpha}}\sin\Delta z \label{eq:two_complex_complex_SLEE}
\end{flalign} where $c := \frac{|\Delta\omega|}{|K|}$. Here, an effective phase-lag parameter is introduced as $\hat{\alpha} :=\alpha - \gamma$. Then, as we demonstrate in the Appendix \ref{appen:complex_frequencies}, the fixed point solutions to Eq.~(\ref{eq:two_general_rescaledtime_SLEE}) with $\gamma \neq 0$ have the same form as Eqs.~(\ref{eq:stable_fixed_comp_SLEE}-\ref{eq:Unstable_fixed_comp_SLEE}) for $\hat{\alpha} \in [0,\frac{\pi}{2})$ and as Eq.~(\ref{eq:fixedpoints_pure_im_SLEE}) for $\hat{\alpha} = \frac{\pi}{2}$, except that the phase-lag parameter $\alpha$ is replaced with $\hat{\alpha} = \alpha-\gamma$. However, their linear stability is not determined by $\hat{\alpha}$, but by the value of $\alpha$ itself since $f'(\Delta z^*) = -e^{\textrm{i}\alpha}\cos \big( \sin^{-1}\Delta z^*\big)$. Therefore, the same dynamical and spectral properties are expected, as classified in the previous sections, with the form of fixed points being determined by the effective phase-lag $\hat{\alpha}$ and their linear stability by $\alpha$: (i) the situation follows that reported in Ref.~\onlinecite{Moritz_thuemler2023} if $\alpha=\gamma$, such that $\hat{\alpha}=0$; (ii) the dynamics of the system for $\hat{\alpha} =\frac{\pi}{2}$ is similar to that discussed in Sec.~\ref{subsec:Two_model_purely_SLEE}. Finally, (iii) for $ \hat{\alpha} \in (0,\frac{\pi}{2})$, the system is expected to have the properties obtained in Sec.~\ref{subsec:Two_model_complex_SLEE}.

\section{\label{sec:finite_ensemble_SLEE}Systems of $N\geq 2$ Complexified Oscillators}

\begin{figure*}[t!]
\includegraphics[width=0.65\textwidth]{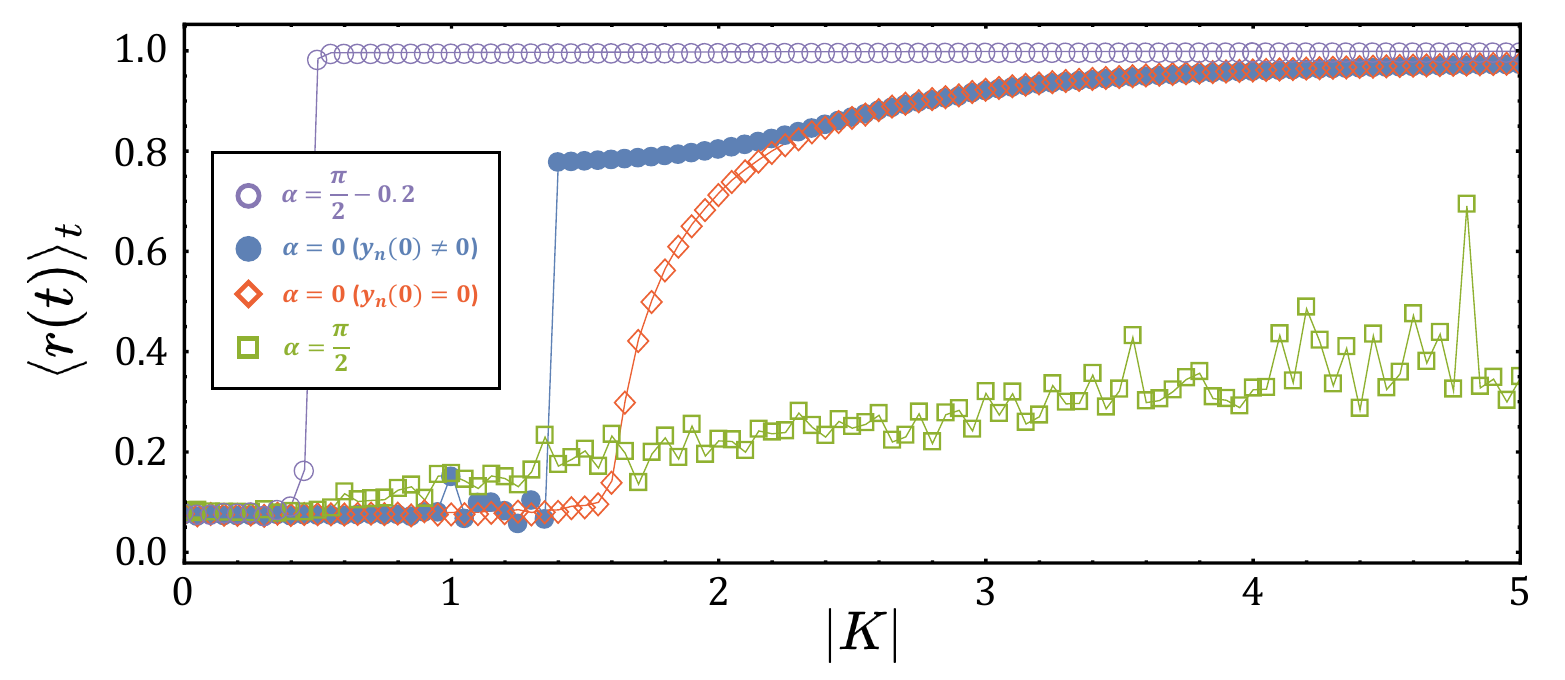}
\caption{\textbf{Large systems of oscillators with complex couplings.} The magnitude of the Kuramoto order parameter $\langle r(t) \rangle_t$ for $N=128$ is depicted as a function of the coupling strength $|K|$: a complex coupling ($\alpha =\frac{\pi}{2}-0.2$, purple open circles), a real coupling ($\alpha=0$, blue solid dots), a real coupling confined within a real-valued invariant Kuramoto manifold ($\alpha=0$, red open rhombuses), and a purely imaginary coupling ($\alpha=\frac{\pi}{2}$, green open squares). Initial conditions are drawn randomly and independently from uniform distributions with $x_n(0) \in [-\pi,\pi]$ and $y_n(0) \in [-10^{-4},10^{-4}]$. } 
\label{Fig:large_N_SLEE}
\end{figure*}

How do the phenomena uncovered for $N=2$ units transfer to larger systems? Here we first consider key aspects for $N\geq2$ coupled complexified Kuramoto oscillators. Rewriting Eq.~(\ref{eq:KM_complexifed_SLEE}) as
\begin{flalign}
    \frac{\diff z_n}{\diff t} &= \omega_n + \frac{K}{N}\sum_{m=1}^N \sin(z_m-z_n) \notag \\
    &= \omega_n + \frac{|K|}{2\textrm{i}N}\sum_{m=1}^{N}\bigg[ e^{\textrm{i}(x_m-x_n+\alpha)}e^{-(y_m-y_n)} \notag \\
    & \hspace{26mm}- e^{-\textrm{i}(x_m-x_n-\alpha)}e^{y_m-y_n} \bigg]
\end{flalign} 
for $n\in[N]$ highlights two general aspects. First, the magnitude $|K|$ of the coupling multiplies each coupling term and thus determines the strength of the coupling, as in real-valued Kuramoto models.  Second, the argument $\arg(K) = \alpha$ takes the role of a phase-lag parameter between two coupled oscillators, akin to the Kuramoto-Sakaguchi model~\cite{sakaguchi1,OMELCHENKO201374}.

Rearranging terms and taking real and imaginary parts, the governing equations read
\begin{align}
\begin{split}\label{eq:complexKM_real_part}
    \frac{\diff }{\diff t} x_n ={}& \Re [\omega_n] +\frac{|K|}{2N}\sum_{m=1}^{N}\bigg[\sin(x_m-x_n+\alpha) e^{-(y_m-y_n)}\\
        & \hspace{26mm} + \sin(x_m-x_n-\alpha) e^{y_m-y_n}\bigg]
\end{split}\\
\begin{split}\label{eq:complexKM_imaginary_part}
    \frac{\diff }{\diff t} y_n={}& \Im [\omega_n ]-\frac{|K|}{2N}\sum_{m=1}^{N}\bigg[\cos(x_m-x_n+\alpha) e^{-(y_m-y_n)}\\
         &\hspace{26mm} - \cos(x_m-x_n-\alpha) e^{y_m-y_n}\bigg]
\end{split}
\end{align} for each $n \in [N]$. 
For example, considering a system of identical complexified oscillators, i.e., $\omega_n=0$ for $n \in [N]$, the phase-lag parameter $\alpha$ delineates the boundary separating regions characterized by linear stability and instability of uniform oscillations, referred to as complete synchronization of complexified Kuramoto oscillators. Here, such instability point occurs at the argument $\alpha=\frac{\pi}{2}$ of the complex coupling $K$. Interestingly, the instability occurs at the same value $\alpha=\frac{\pi}{2}$ in a system of real-valued Kuramoto-Sakaguchi oscillators.

To explore the collective behavior in a system of complexified Kuramoto oscillators with distributed natural frequencies, we deterministically select their real parts from a Gaussian distribution $g(\omega) = \frac{1}{\sigma \sqrt{2\pi}} e^{-\frac{\omega^2}{2 \sigma^2}} $ with zero mean and variances $\sigma$ that we here take to be $\sigma =1$. Specifically, we choose a set of natural frequencies according to \textit{inverse transform sampling}, i.e., with $\Re[\omega_n] = \sqrt{2}\sigma \textrm{erf}^{-1}\big( \frac{2n-1-N}{N} \big)$ for $n \in [N]$, where $\textrm{erf}$ is the error function. This way we choose a finite number of frequencies equally spaced in probabilities, i.e. as arguments of an inverse cumulative distribution function of $g(\omega)$, thereby enforcing  $\sum_{n=1}^N \Re[\omega_n]=0$ exactly for finite $N$. For the example study below, we pick all imaginary parts to be zero, $\Im[\omega_n]=0$ for $n \in [N]$, and select real parts $\Re[\omega_n]$ as above.

Figure ~\ref{Fig:large_N_SLEE} depicts time-averaged Kuramoto order parameters~(\ref{eq:Kuramoto_order_SLEE}), directly measured from numerical simulations, as a function of the modulus $|K|$ of the coupling $K = |K|e^{\textrm{i}\alpha}$ in four different settings. First, for real coupling (with $\alpha = 0$) and   initial conditions with zero imaginary parts ($y_n(0)=0$ for all $n \in [N]$), a transition occurs as in the real-variable Kuramoto model. With increasing $|K|$, a partially locked state emerges from an incoherent state (red open diamonds). Indeed, the oscillators are constrained onto the real invariant manifold \eqref{eq:invariant_KM_manifold_SLEE}, so the dynamics exactly resembles that of the original,  real-valued Kuramoto model, cf. also Fig.~\ref{Fig:real_KM_SLEE}a. In contrast, the same system started from initial conditions with non-zero imaginary parts ($y_n(0)\neq0$ for all $n \in [N]$) displays markedly different features (cf. Ref.~\onlinecite{Moritz_thuemler2023}): The Kuramoto order parameter discontinuously jumps from indicating an incoherent state coexisting with unstable complex-locked states to a stable complex-locked state, i.e., a stable fixed point in complex vector space defined in Ref.~\onlinecite{Moritz_thuemler2023}. 

For generic complex coupling $\alpha=\frac{\pi}{2}-0.2$ (purple circles in Fig.~\ref{Fig:large_N_SLEE}) the system exhibits a discontinuous synchronization transition at substantially smaller $|K|$, where the majority of oscillators become closely locked and $r$ is close to one. Past the transition, pronounced synchrony emerges with a distribution variance in the real parts $x_n$ on the order of $10^{-3}$ and non-zero imaginary parts. Finally, for a purely imaginary coupling with $\alpha=\frac{\pi}{2}$, the oscillators are incoherently drifting, yielding a low order parameter relative to other scenarios even at substantially large $|K|$  (green open squares). These findings qualitatively align with those for the two-oscillator systems discussed in Sec.~\ref{subsec:real_omega_SLEE}, where the systems fail to exhibit any classical locked state and are asynchronous for purely imaginary coupling, whereas oscillators subject to a generic complex coupling become locked at a stable fixed point and exhibit a substantial degree of synchrony.

\section{\label{sec:conclusions} Conclusions and Open Questions}

We have extended the Kuramoto model with analytically continued state variables \cite{Moritz_thuemler2023} by also complexifying the parameters, and analyzed their collective dynamics.

Our results illustrate a broad variety of novel ordering phenomena in time, including indications for discontinuous transitions, early transitions to strong synchrony and the persistence of asynchronous dynamics even at arbitrarily large $K$. Simultaneously, as previous works on complexified Kuramoto models indicate \cite{Moritz_thuemler2023,Roberts2008,Muller2021,Seung_Yeal2012}, studying models on fully complexified state spaces and their associated state space topologies offer a more integrated perspective. For instance, stable, unstable and neutrally stable fixed points in the complex domain control collective system dynamics. As these do not exist in the real state space analogs, transient dynamics emerges in real models thereby hindering the analysis of transitions between more and less ordered states.

More specifically, we have presented a range of results:
    For $N=2$ coupled units and purely imaginary coupling, we discovered motions of \textit{libration} around fixed points of in-phase and antiphase states which are respectively local minima and maxima of a conserved quantity, and a ceaseless \textit{rotation} along the cylindrical axis with an infinite rotation number $q=\infty$. No attractive locked state emerges even for arbitrarily large $|K|$.
    In contrast, for generic complex coupling, trajectories rotate a few times, exhibit finite rotation numbers $q<\infty$, and stop at a stabilized fixed point. We have analytically determined the rotation number and derived its asymptotic scaling as $\alpha\rightarrow \pi/2$ from below, where the coupling becomes purely imaginary. 
    For small $\beta=\pi/2-\alpha$, synchrony stabilizes: Slightly turning on the real part of a coupling, the neutrally stable fixed points of in-phase synchrony become stabilized to first order in $\beta$.
    Larger, finite-size systems of $N>2$ units similarly exhibit a classically locked state for purely imaginary coupling whereas a \textit{discontinuous transition} emerges to a locked state for a generic complex coupling.
%
The analyses above catalyze a range of future research.
\begin{enumerate}
    \item The \textit{thermodynamic limit}. Does the critical point at which a complex locked state becomes stable indicate transition properties in the real-variable system, perhaps indirectly? An appropriately generalized order parameter and self-consistency equation or dimension reduction methods like Ott-Antonsen~\cite{OA1,OA2,OA3} or Watanabe-Strogatz~\cite{mobius_WS,WS-original2,pikovsky_WS2} approaches might help address such questions. For instance, related recent work has demonstrated that for identical complexified oscillators, the dimension reduction method is applicable using the complex Riccati equation~\cite{cestnik2024,lohe_riccati}. However, it remains unclear how these results generalize to systems with heterogeneous natural frequencies.
    \item What is the collective dynamics of \textit{spatially extended} systems of complexified oscillators, for instance on a ring or other, higher-dimensional geometries such as a two-dimensional plane, sphere, or torus? Spatiotemporal dynamics such as a traveling wave or chimeras might yield starting points for novel behavior. Specific questions may cover, importantly, the role of the imaginary parts. 
   
    Are complexified systems of oscillators capable of describing chemical reactions or biological patterns in real systems, perhaps similar in spirit to the complex Ginzburg-Landau equations~\cite{cgle_krischer,world-cgle}?
    \item What is the dynamics of complexified oscillators on \textit{non-trivial network topologies}? Specifically, how does the network topology affect the state of synchrony and the transition(s) towards it? The main results may find a broad range of applications, from understanding generic bifurcations in toy-model networks to real networks, for instance addressing how power grids lose stability.
    \item What is the dynamics of analytically continued phase oscillator networks and \textit{coupled dynamical systems} models in general? Are there typical, system-overarching features? If, so which are they and which features are system-dependent? Suitable candidates for initial studies include the Winfree model~\cite{Seung_Yeal2021}, the theta neuron model~\cite{cestnik2024,Ermentrout_2008}, the (second order) Kuramoto model with inertia that models power grid dynamics and also features intriguing collective dynamics, van der Pol oscillators and other models. 
\end{enumerate}
    In summary, complexification by analytical continuation of state variables\cite{Moritz_thuemler2023} and complexification of parameters may serve as a general analysis tool for dynamical systems. It seems particularly promising as the complexified systems may exhibit fixed points (e.g. complex locked states) that are mathematically accessible whereas the original real-variable systems only exhibit transients that require intricate numerical simulations. Complexification may thus enable a deeper understanding of the respective original systems as well as unveil a number of collective phenomena worth investigating in their own right.

\section*{Nonlinear Dynamics with David K. Campbell}
David Campbell, who is approaching his 80th birthday in 2024, has been a pioneer of nonlinear dynamics, has co-established the field and impacted many paths of individual researchers within it. David has repeatedly taken motivation from physics and in turn demonstrated diverse valuable applications in physics and beyond. By often leaving well-trodden paths and pointing to novel perspectives, he has left his personal mark in the field. His groundbreaking contributions to nonlinear phenomena include, for instance, revealing conditions and mechanisms underlying solitary waves and discrete breathers that emerged from more traditional subfields such as solid-state physics. 
Through many decades of work and inspiration, he has created an enduring legacy that overarches the full range from observed nonlinear phenomena in nature to advancing mathematical tools for their analysis. Thank you David, for advice, for co-creating the field, and for showing the path forward. May you have a truly happy birthday and a happy nonlinear time to come.

\section*{Data Availability Statement}

The data that support the findings of this study are available from the first author upon reasonable request.

\section*{Author Contributions}
\textbf{Seungjae Lee}: Conceptualization (equal); Formal analysis (lead); Software (lead); Writing – original draft (lead); Writing – review \& editing (lead). \textbf{Lucas Braun}: Formal analysis (supporting);
Software (supporting); Writing – review \& editing (supporting). \textbf{Frieder B\"onisch}: Formal analysis (supporting); Software (supporting); Writing – review \& editing (supporting). \textbf{Malte Schr\"oder}: Formal analysis (supporting); Writing – review \& editing (supporting). \textbf{Moritz Th\"umler}: Conceptualization (equal); Formal analysis (supporting); Writing – review \& editing (supporting). \textbf{Marc Timme}: Conceptualization (equal); Formal analysis (lead); Supervision (lead); Writing – original draft (lead); Writing – review \& editing (lead).

\begin{acknowledgments}
Partially supported by German Federal Ministry for Education and Research (BMBF) under grant number 03SF0769 (ResiNet) as well as the German National Science Foundation (Deutsche Forschungsgemeinschaft DFG) under grant number DFG/TI 629/13-1 and under Germany´s Excellence Strategy – EXC-2068 – 390729961, through the Cluster of Excellence \textit{Physics of Life}.
\end{acknowledgments}

\appendix

\section{\label{appen:complex_frequencies}Complex Intrinsic Frequencies}

In this Appendix, we provide details of linear stability analysis of fixed point solutions to Eq.~(\ref{eq:two_general_rescaledtime_SLEE}) discussed in Sec.~\ref{subsec:general_two_SLEE}: (i) For $\hat{\alpha}=0$, i.e., $\alpha=\gamma$, the fixed points read $\Delta z^* = \sin^{-1}c$. Here, we follow the same argument reported in Ref.~\onlinecite{Moritz_thuemler2023}. First, two representative fixed points for $c<1$ read $\Delta z^*_{0} = \sin^{-1}c$ and $\Delta z^*_{1} = \pi-\sin^{-1}c$. For these fixed points, eigenvalues of the Jacobian matrix of Eq.~(\ref{eq:two_general_rescaledtime_SLEE}) are given as $f'(\Delta z^*_{0,1}) = \mp e^{\textrm{i}\alpha}\sqrt{1-c^2}$ together with their complex conjugates. Hence, $\Delta z^*_0$ ($\Delta z^*_1$) is stable (unstable) for $\alpha<\frac{\pi}{2}$ whereas $\Delta z^*_0$ ($\Delta z^*_1$) becomes unstable (stable) for $\alpha>\frac{\pi}{2}$. Secondly, for $c=1$, a degenerate semi-stable fixed point occurs at $\Delta z^*=\pi/2$ with a tilted center manifold with respect to the real axis. Note that the center manifold of the semi-stable fixed point for $\gamma=0$ and $\alpha=0$ is on the real axis. See Fig. 2b in Ref.~\onlinecite{Moritz_thuemler2023}. Thirdly, for $c>1$, the fixed points are given as $\Delta z^*_{0,1} = \frac{\pi}{2} \pm \textrm{i}\log (c+\sqrt{c^2-1})$ and eigenvalues of the Jacobian matrix to Eq.~(\ref{eq:two_general_rescaledtime_SLEE}) for these fixed points are $f'(\Delta z^*_{0,1}) = \pm \textrm{i}e^{\textrm{i}\alpha}\sqrt{c^2-1}$, which confirms that $\Delta z^*_0$ ($\Delta z^*_1$) is stable (unstable) for $\sin\alpha >0$ whereas $\Delta z^*_0$ ($\Delta z^*_1$) becomes unstable (stable) for $\sin\alpha <0$. Hence, a bifurcation occurs at $\alpha = \pi$. (ii) In the case of $\hat{\alpha}=\pi/2$, the system is investigated as discussed in Sec.~\ref{subsec:Two_model_purely_SLEE}. It has fixed points that read $\Delta z^*_0 = -\textrm{i}\sinh^{-1}c$ and $\Delta z^*_0 = \pi + \textrm{i}\sinh^{-1}c$ with $f'(\Delta z^*_{0,1}) = \mp e^{\textrm{i}\alpha} \sqrt{1+c^2} $, confirming the former (the latter) is stable (unstable) for $\cos\alpha > 0$ and becomes unstable (stable) for $\cos\alpha < 0$. (iii) For $\hat{\alpha} \in (0,\frac{\pi}{2})$, the fixed points are given as Eqs.~(\ref{eq:stable_fixed_comp_SLEE}-\ref{eq:Unstable_fixed_comp_SLEE}) replacing $\alpha$ by $\hat{\alpha}$ and are explored as in Sec.~\ref{subsec:Two_model_complex_SLEE}, yet the linear stability determined by the value of $\alpha$.

\bibliography{bibliography}

\end{document}